\documentclass[rnote]{aa} 
\usepackage{graphicx}
\usepackage{txfonts}
\usepackage{longtable}
\usepackage{xtab}
\usepackage{rotating}
\usepackage{lscape}
\usepackage{natbib}
\usepackage{color}
\bibpunct{(}{)}{;}{a}{}{,} 
\begin{document}
   \title{Coordinates and 2MASS and OGLE identifications for all stars
in Arp's 1965 finding chart for Baade's Window}
 \titlerunning{Coordinates for Arp's Baade's window stars}

   \author{R.P. Church
          \inst{1}
	  J.A. Johnson
	  \inst{2,1}\thanks{J.A. Johnson is a guest professor at Lund Observatory}
	  S. Feltzing
	  \inst{1}
          }

   \institute{Lund Observatory, Department of Astronomy and Theoretical
Physics, Box 43, SE-22100, Lund, Sweden\\
              \email{ross, sofia@astro.lu.se}
         \and
             Department of Astronomy, Ohio State University, 140 W. 18th Avenue, Columbus, OH 43210, USA\\
             \email{jaj@astronomy.ohio-state.edu}
             }

   \date{Received ; accepted }

 
  \abstract
   {}
  {We seek to provide 2MASS and OGLE identifications and coordinates for all
    stars in the finding chart published by Arp\,(1965).
    This chart covers the low extinction area around
    NGC\,6522, also known as Baade's window, at coordinates  $(l,b)=(1.02,-3.92)$.}
   {A cross correlation, using numerical techniques, was performed between 
a scan of the original finding chart from Arp (1965) and 2MASS and OGLE-II
images and stellar coordinates.}
   {We provide coordinates for all stars in Arp's finding chart and 2MASS
and OGLE identifications wherever possible. Two identifications in quadrant II
do not appear in the original finding chart.}
   {}


   \maketitle
%

\section{Introduction}

The region towards the Galactic Bulge situated at Galactic coordinates
$(l,b)=(1,-4)$ known as Baade's window is a much studied area as it
provides a low extinction region where we are able to look deeper into
the Galactic Bulge than in most other regions. This region is centred
on a globular cluster, NGC\,6522. Halton~\citet{arp1965} provided a much-used
finding chart for the immediate region around the cluster. This article also
provided broad band photometry for all the labelled stars. This finding chart has
proved important especially for the identification of stars suitable for high
resolution spectroscopy
\citep[e.g.,][]{1994ApJS...91..749M,fulbright2006,ryde2009}. However,
\cite{arp1965} did not publish coordinates for the stars, and very few
of the spectroscopic studies using his finding chart have done so.
Indeed, only 37 of the stars in \cite{arp1965}
have SIMBAD entries (according to ADS 2011-02-15).  

In the era of large surveys, such as Sloan Digital Sky Survey (SDSS)
and The Two Micron All Sky Survey
\citep[2MASS,][]{2006AJ....131.1163S}, old numbering schemes for
relatively small patches of the sky may appear of less
relevance. However, for correct comparison with the high resolution
studies it is important that we know which star has which
identification number in Arp's catalogue and which 2MASS and 
Optical Gravitational Lensing Experiment (OGLE) identifications it
has. We suspect that others have carried out the same painful exercise
as we have done a few times and truly cross-identified some of, e.g.,
the \cite{fulbright2006} stars with their correct 2MASS
identification. We are, however, not aware of any such list available.

We therefore decided to provide a complete cross identification
between the Arp numbers and 2MASS and OGLE identifications and provide
coordinates for all stars on the finding chart in \citet{arp1965}.

\section{Method of cross identification}
\label{sect:method}

To obtain the co-ordinates of the stars in Arp's sample, we took a scan
of figure~1 of \citet{arp1965} at 600\,dpi using an office photocopier.  The
halftone pattern produced by the printing process was removed by iteratively
applying a simple algorithm in which white pixels with predominantly black
neighbours were coloured black and vice-versa.  Having done this we identified
all contiguous regions of black points and analysed each in turn, in an attempt
to determine whether they were stars or Arp's annotations.  The smallest regions
($<50\,{\rm pixels}$) were disregarded, and regions with a high aspect ratio
were marked as lines.  Similarly, thin sections of regions were marked as lines
protruding from stellar images.  The remaining regions contained predominantly
stars and numeric digits.  We selected stars as regions that had relatively high
surface density; i.e. where $>60\%$ of the rectangular area encompassing the
region was black pixels.  We found this to work well as stars tended to be solid
blocks of pixels, whereas numbers tend to contain empty regions.  Empirically we
also found it useful to repeat this test with the outer 20\% of the region
removed and using a higher threshold of 66\%.  Unusually large regions
($>750\,{\rm pixels}$) were marked as stars, as well as small regions
($<300\,{\rm pixels}$, but not including the very small regions disregarded
above).  Whilst a more sophisticated probabilistic model might have been more
accurate, these empirical rules served to separate stars from annotation marks
with an accuracy of about 90\%.

For each identified stellar image we calculated the centroid.  We
checked the location of the centroid of each identified star by eye
and adjusted those where the algorithm had performed badly.  In many
cases a pair of stellar sources had become blended into a single
region; in these cases the centroids of the two stars were extracted
using the K-means++ algorithm \citep{Arthur2007}.  We then matched
each stellar object with its number -- if any -- in Arp's catalogue by
hand.

In order to cross-identify Arp's sources with 2MASS sources we used
the sample of \citet{Blanco86} and positions from \citet{Glass02} to
produce an initial mapping between pixel co-ordinates in our scanned
image and sky positions.  Right ascension and declination were fit as
linear combinations of $x$ and $y$ in pixel space, with a separate fit
being made for each of Arp's quadrants.  We used this fit to
identify, for each star, the closest 2MASS source.  Where the closest
source was within 2 arcseconds, this was taken to be a tentative
match, and the fit repeated using the catalogued positions of these
stars.  We checked each match by eye to see whether it appeared to be
genuine; in some cases a single 2MASS source was seen to be a blending
of two sources in Arp's catalogue.  Some of Arp's stars were found to
have no corresponding source in 2MASS.  For these sources, and sources
which appeared to be blended into one 2MASS source we present
co-ordinates derived from our fitting; otherwise the co-ordinates
presented are those of the corresponding 2MASS source. That not 
all stars in Arp's finding chart have 2MASS counterparts is a
natural consequence of the redder filters used in 2MASS and the
fact that the two surveys reach different depths. 

Finally we repeated the fitting process to match stars in the OGLE
catalogue, where the cross-over was more complete.  All OGLE sources
identifications are in the BUL SC45 field and we used an $I$-band
template image from OGLE-II for the cross-correlation \citep{2002AcA....52..217U}.

Identification numbers II-7 and 39 appear to not have any counterparts
on the original finding chart in \citet{arp1965}.

\section{Table of coordinates, identifiers, and photometry}

Table\,1 presents  our results for Arp's quadrants I, II, III and
IV\footnote{This table is available in the online version of this paper and at
CDS.}. Here we give a brief description of the contents of Table\,1.

\begin{description}
\item[{\sl Id number}] The first two columns give the identifications from
the finding chart in \citet{arp1965}. The first column gives them as Arabic
digits; the first digit indicates the quadrant. The second column gives
the classical notation used in \citet{arp1965} with quadrants indicated
by roman numerals.

\item{{\sl Coordinates}} Coordinates are from 2MASS whenever
  available, for the remaining stars we give the coordinates as
  calculated in Sect.\,\ref{sect:method}. The coordinates are in
  J2000 and are on the system set by 2MASS.

\item{{\sl 2MASS (incl. blends) and OGLE identifications}} We list the 
2MASS and OGLE identifications and for 2MASS we also indicate whether 
or not the identification was a blend (see  Sect.\,\ref{sect:method}).

\item{{\sl Photometry}} For convenience we include the photometry from
  2MASS ($JHK_S$) and OGLE-II \citep[$VI$
  photometry][]{2002AcA....52..217U} in the table. We have included
  certain flags regarding the quality of the 2MASS photometry. The
  description of the meaning of the relevant flags are given in
  Appendix\,\ref{sect:app}. We include the errors in the photometry
for both 2MASS and OGLE data. 

\end{description}

\section{Summary}

We present coordinates for all stars identified in the finding chart
by \citet{arp1965} of the globular clusters NGC\,6522, also known as
Baade's window. This is a much studied low extinction area situated close to
$(l,b)=(1,-4)$. The coordinates are on the 2MASS system. 

As part of the process we have produced, in a semi-automated fashion,
a match between the sources in \citet{arp1965} and the 2MASS and OGLE
catalogues.

\begin{acknowledgements}
  RPC is funded by a Marie-Curie Intra-European Fellowship, grant No.~252431,
  under the European Commission's FP7 framework. JAJ acknowledges a
  guest professorship provided by the Faculty of Science, Lund
  University.  SF was partly supported by grant No. 2008-4095 from The
  Swedish Research Council.

  This research has made use of the SIMBAD database, operated at CDS,
  Strasbourg, France. 

  This publication makes use of data products from the Two
  Micron All Sky Survey, which is a joint project of the University of
  Massachusetts and the Infrared Processing and Analysis
  Center/California Institute of Technology, funded by the National
  Aeronautics and Space Administration and the National Science
  Foundation. 

\end{acknowledgements}

\bibliographystyle{aa}
\bibliography{references}

\begin{appendix}

\section{Flags for 2MASS photometry}
\label{sect:app}

\noindent
A = Detections in any brightness regime where valid measurements were
made with [jhk]snr$>$10 AND [jhk]cmsig$<$0.10857.

\noindent
B = Detections in any brightness regime where valid measurements were
made with [jhk]snr$>$7 AND [jhk]cmsig$<$0.15510.

\noindent
C = Detections in any brightness regime where valid measurements were made  with [jhk]snr$>$5 AND [jhk]cmsig$<$0.21714.

\noindent
{\sl D} = Detections in any brightness regime where valid measurements were
made with no [jhk]snr OR [jhk]cmsig requirement.

\noindent
{\sl E} = This category includes detections where the goodness-of-fit
quality of the profile-fit photometry was very poor, or detections
where psf fit photometry did not converge and an aperture magnitude is
reported, or detections where the number of frames was too small in
relation to the number of frames in which a detection was
geometrically possible.

\noindent
{\sl U} = Upper limit on magnitude. Source is not detected in this band, or
it is detected, but not resolved in a consistent fashion with other
bands.

\end{appendix}

\Online

\begin{landscape}
\tablecaption{Cross-identification and photometry of all stellar sources}
\tablehead{\hline\hline%
\multicolumn{2}{c}{Arp Id} & Ra    &  Dec  &  Ra & Dec & 2MASS Id & Bld & $J$ & $e_J$ & $H$ & $e_H$ & $K$ & $e_K$ &   Qflag & OGLE Id & $V$ & $V-I$ & $I$   & $e_V$ & $e_I$ \\
                &          & h m s & d m s & deg & deg &                       &     & mag &  mag  &  mag&   mag & mag &  mag  &        &          & mag & mag   & mag   &  mag  &  mag  \\ \hline}
\tabletail{\hline}
\scriptsize
\begin{xtabular}{lllllllllllllllllllll}
\input{table.txt}
\hline
\end{xtabular}
\end{landscape}

\end{document}